\documentclass[aps,prb,twocolumn]{revtex4}

\usepackage[]{graphicx}

\begin{document}

\title{Large-area, wide-angle, spectrally selective plasmonic absorber}









\author{Chihhui Wu, Burton Neuner III, and Gennady Shvets}

\email{gena@physics.utexas.edu}




\affiliation{Department of Physics, The University of Texas at Austin, Austin, TX 78712}

\author{Jeremy John, Andrew Milder, Byron Zollars, and Steve Savoy}
\affiliation{Nanohmics, Inc. 6201 E. Oltorf, Suite 400, Austin, TX 78741}




\date{\today}

\begin{abstract}
A simple metamaterial-based wide-angle plasmonic absorber is introduced, fabricated, and
experimentally characterized using angle-resolved infrared spectroscopy. The metamaterials are
prepared by nano-imprint lithography, an attractive low-cost technology for making large-area
samples. The matching of the metamaterial's impedance to that of vacuum is responsible for the
observed spectrally selective ``perfect" absorption of infrared light. The impedance is
theoretically calculated in the single-resonance approximation, and the responsible resonance is
identified as a short-range surface plasmon. The spectral position of the absorption peak (which is
as high as $95\%$) is experimentally shown to be controlled by the metamaterial's dimensions. The
persistence of ``perfect" absorption with variable metamaterial parameters is theoretically
explained. The wide-angle nature of the absorber can be utilized for sub-diffraction-scale infrared
pixels exhibiting spectrally selective absorption/emissivity.
\end{abstract}


\maketitle 



\section{Introduction}

The field of electromagnetic metamaterials (MMs) has been rapidly developing in recent years. New
composite materials with sub-wavelength size and exotic electromagnetic properties generally
unattainable in nature~\cite{veselago_spu68,smith_prl00,pendry_prl00,schurig_science06} are being
designed and produced for many applications such as perfect lenses~\cite{pendry_prl00}, cloaking
devices~\cite{schurig_science06}, sub-wavelength transmission lines and
resonators~\cite{alu_ieee04}, and agile antennas~\cite{chen_np08}. Another important application of
MMs is the development of spectrally selective ``perfect" absorbers~\cite{padilla_absorber_08}
(near-unity peak absorptivity). Such absorbers can be used for developing sensitive detectors for a
variety of security-related applications, as well as narrow-band thermal emitters for
thermophotovoltaic (TPV)~\cite{coutts_rsev99,laroche_jap06} applications. In photovoltaic (PV)
applications, ultra-thin MM absorbers can be applied on surfaces of thin-film solar cells to reduce
reflectivity~\cite{tvingstedt_apl07}, thereby increasing the external quantum efficiency.

Miniaturization of these devices is highly desirable and can be achieved by making the
metamaterial's unit cell strongly sub-wavelength~\cite{avitzour_prb09, wu_spie08, soukoulis_prb09,
padilla_apl10, giessen_nl10}. Strong confinement of electromagnetic fields to sub-wavelength
regions of the resonant metamaterials results in spectrally selective absorption. An added benefit
of the sub-wavelength unit cell is the wide-angle response of the metamaterials. The wide-angle
response is important for making hyper-spectral focal plane arrays (FPA) comprised of ultra-small
detector pixels. Angular directivity can also be detrimental for TPV applications because it
effectively broadens the emission spectrum. While miniaturizing the unit cell for microwave/THz
applications can be done using traditional MM approaches---such as making split-ring
resonators~\cite{tao_prb08}---fabrication challenges make such approaches impractical for optical
MMs. Therefore, plasmonic resonances of much simpler metallic structures~\cite{urzh_shvets_ssc08}
must be utilized to reduce the unit cell's size in optical MMs. Examples of such structures used in
MM-based optical absorbers include metal strips~\cite{wu_spie08, soukoulis_prb09} and
patches~\cite{padilla_apl10, giessen_nl10} separated by a thin dielectric layer from a metallic
ground plate. While the wide-angle absorption of these structures has been theoretically
demonstrated~\cite{wu_spie08, soukoulis_prb09, padilla_apl10, giessen_nl10}, no experimental
evidence presently exists.

Here we report on the fabrication and experimental demonstration of wide-angle, spectrally
selective plasmonic surfaces exhibiting near-unity absorption of infrared radiation. The fabricated
structure schematically shown in Fig.1(a) is comprised of plasmonic strips separated from the
plasmonic ground plate by an ultra-thin ($< \lambda/50$) dielectric spacer. Metamaterials with
various unit cell dimensions and spacer materials were prepared using the method of ultraviolet
nano-imprint lithography (NIL)~\cite{johnson_spie03}. The NIL process provides a means for
large-area replication of the device pattern after an initial electron beam lithographic process is
used to generate the reticle. This MM absorber encompasses the following advantageous
characteristics: extremely high absorption approaching $100\%$ over a wide range of incident
angles, tunability of the absorption peak frequency demonstrated by changing the unit cell's
dimensions, robustness against structure imperfections, sub-wavelength ($< 1~\mu$m) unit cell
sizes, strong field confinement within the MM structure, and large MM area ($> 1$~mm$^2$ per MM
pattern).

The rest of the paper is organized as follows. In Section~\ref{sec:theory}, we describe theoretical
tools used in calculating the effective impedance of a resonant plasmonic surface and introduce the
single-resonance impedance model. Using eigenvalue simulations, we compute the complex impedance of
the plasmonic absorber and explain the phenomenon of ``perfect'' absorption in terms of impedance
matching of the absorber to that of vacuum. In Section~\ref{sec:fabrication}, we describe the
fabrication of the absorber using NIL. In Section~\ref{sec:spectroscopy}, 
we demonstrate high absorptivity and spectral tunability of the absorber both experimentally and theoretically. An external 
beamline based on FTIR spectroscopy was used to verify the wide-angle absorptivity predicted by numerical 
calculations. Implications of the wide-angle absorptivity are discussed in Section~\ref{sec:small_pixel}. The 
surface mode responsible for the high absorptivity is shown to be highly localized and has a propagation 
length shorter than the structure periodicity. Unit cells of the absorber function almost independently 
and have little cross-talk.

%

\section{Theoretical Description of the Frequency-Selective
Plasmonic Absorber}\label{sec:theory}

\begin{figure}[t]
 \centering
 \includegraphics[width=.4\textwidth]{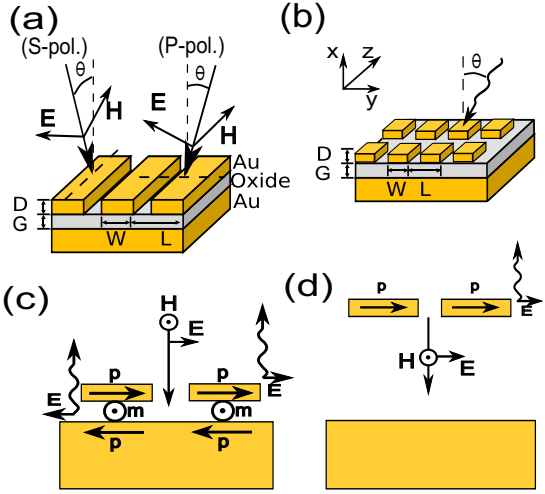} \
 \caption{Schematic of the plasmonic absorber with (a) parallel
 strips and (b) square patches placed above the ground plate.
 (c), (d) Reduced
 radiative loss rate due to destructive interference between ground
 plate reflection and magnetic dipole radiation. (c) Strong
 magnetic dipoles are induced when $G$ is small; (d) negligible
 magnetic dipoles are induced when $G$ is large.}\label{fig:schematic}
\end{figure}

In the following we concentrate on the specific geometry of a resonant plasmonic absorber
schematically shown in Fig.~\ref{fig:schematic}(a,b), which is comprised of a periodic array of
either plasmonic strips or square patches separated from a thick (non-transparent) plasmonic
mirror (further referred to as the ground plate following the standard microwave terminology). As
shown in Section~\ref{sec:spectroscopy}, the spectral response of the square-patch structure shown
in Fig.~\ref{fig:schematic}(b) is qualitatively (and even quantitatively) similar to that of the
one-dimensional structure. In this section we concentrate on the one-dimensional array of plasmonic
strips~\cite{wu_spie08} shown in Fig.~\ref{fig:schematic}(a), where all dimensions of the plasmonic
absorber are defined. The structure is assumed uniform in the $z$-direction and repeating in the
$y$-direction with periodicity $L$. Below we develop the theoretical formalism for computing the
absorption of the nano-structured plasmonic surface based on the eigenvalue/eigenmode analysis. For
simplicity, we restrict the incident wave to be P-polarized as shown in
Fig.~\ref{fig:schematic}(a), i.e., the incident wavenumber and the electric field lie in the
$x$-$y$ plane, whereas the magnetic field is aligned in the $z$-direction.

In earlier reports on MM absorbers, the perfect absorption was explained in terms of matching the
bulk metamaterial's impedance $z=\sqrt{\mu_{\it eff}/\epsilon_{\it eff}}$ to that of
vacuum~\cite{yoav_prb09,padilla_prl10,padilla_prl08,padilla_prb09}, where $\epsilon_{\it eff}$ and
$\mu_{\it eff}$ are the effective permittivity and permeability of the bulk multi-layer (i.e.,
stacked in the vertical $z$-direction) metamaterial. Typically, these effective constitutive
parameters are obtained from the scattering matrices of either a single layer or multiple layers of
the MMs~\cite{smith_prb02} under the assumption of homogeneous media. To satisfy the condition of
matched impedances, the condition of $\epsilon_{\it eff}=\mu_{\it eff}$ (which are both, in
general, complex numbers) is achieved~\cite{yoav_prb09} by manipulating the spectral positions and
strengths of the electric and magnetic resonances of the MMs. However, this viewpoint is not
appropriate for MMs using a thick ground plate because the S-matrix is not fully defined when
transmittance vanishes. Additionally, the ambiguity in the MM thickness also poses a problem in
calculating $\epsilon_{\it eff}$ and $\mu_{\it eff}$.
\begin{figure}[t]
 \centering
 \includegraphics[width=.45\textwidth]{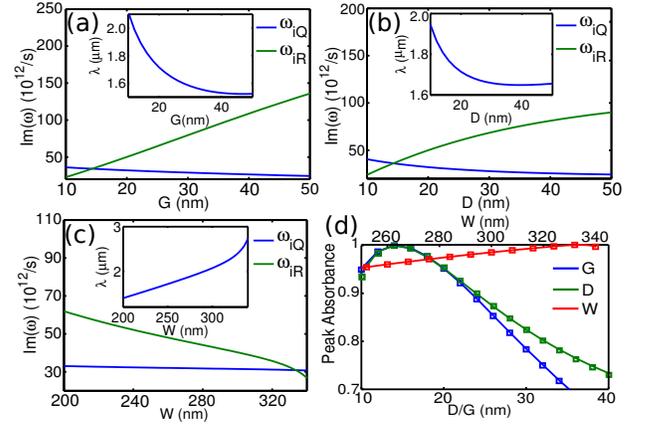} \
 \caption{Resistive ($\omega_{iQ}$) and radiative ($\omega_{iR}$)
 loss rates of the eigenmodes as functions of (a) the dielectric
 gap size ($G$), (b) metal strip thickness ($D$), and (c) metal
 strip width ($W$). The parameters are varied around $L$=350~nm,
$W$=250~nm, $D$=20~nm, and $G$=20~nm. Equal resistive and
radiative loss rates correspond to critical coupling and perfect
absorption. Insets: dependence of the resonant wavelengths of the
corresponding eigenmodes. (d) Peak absorbance calculated from
eigenmode (markers) and driven simulation (solid lines), with
the parameters in the same range as
(a-c).}\label{fig:cross-section}
\end{figure}

Here we explain the phenomenon of perfect absorption in terms of critical coupling to the MM
surface. This perspective is more adequate for our situation with blocked transmission. It provides
a guideline on how the structure should be designed, and also reveals the persistence of high
absorbance over structural variation. Critical coupling occurs when a leaky eigenmode of the
structure has equal resistive and radiative losses. For a critically coupled MM, an incoming field
excites the eigenmode through radiative coupling, and the incoming energy transforms into resistive
loss without generating reflection. Therefore, a MM absorber with blocked transmission is analogous
to the single resonator coupled to a single input waveguide. This problem has been well-studied~\cite{haus_1984}, 
and the reflection coefficient $r$ can be expressed as a function of the
resonant frequency, $\omega_0$, the radiative damping/coupling rate, $\omega_{iR}$, and the resistive damping rate, 
$\omega_{iQ}$, of the resonator. The impedance of the MM surface defined as $z=(1+r)/(1-r)$ is given by:

\begin{eqnarray}
&&z=\frac{\omega_{iR}}{i(\omega-\omega_0)+\omega_{iQ}}.
\label{eq:impedance}
\end{eqnarray}
From Eq.~(\ref{eq:impedance}), the critical
coupling condition $z=1$ is satisfied when $\omega_{iR}=\omega_{iQ}$. Because the ground plate
transmits no light, the absorptivity of the structure $A(\omega)$, which is determined by its
reflectivity according to $A\equiv 1-|r|^2$, reaches unity at $\omega=\omega_0$. Note that
$r(\hat{\omega})$ diverges at the complex frequency $\hat{\omega}$ if $z(\hat{\omega})=-1$, or
$\hat{\omega} = \omega_0 - i \omega_i$, where $\omega_i \equiv \omega_{iR} + \omega_{iQ}$. Because
the divergence of the reflectivity corresponds to the eigenmode of the ``leaky" resonator (or, in
our case, leaky plasmonic absorber), we can calculate $\hat{\omega}$ by computing the complex
eigenfrequency of the absorber. The corresponding plasmonic eigenmode is further used to break up
the total decay rate $\omega_i$ into the Ohmic (resistive) and radiative parts. A similar
approach~\cite{neuner_OL09} was recently used to describe critical coupling to surface polaritons
excited in the Otto configuration.
\begin{figure}[t]
 \centering
 \includegraphics[width=.45\textwidth]{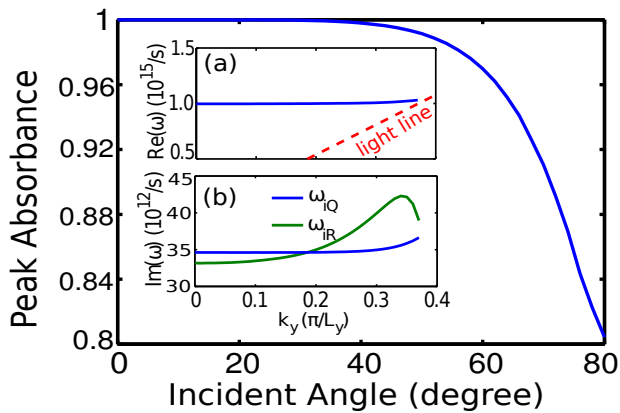} \
 \caption{Angular dependence of the plasmonic
 resonance responsible for ``perfect" absorption. (a) Real and (b)
 imaginary parts of the
 eigenfrequency as a function of the wavenumber $k_y=\sin{\theta}
 \omega/c$ in the periodicity direction. The imaginary part is
 separated into radiative and Ohmic loss rates. (c) Peak
 absorbance remains above $80\%$ for all incidence angles
 $\theta$.}\label{fig:angular_theory}
\end{figure}
The calculation was performed using the finite element method (FEM) software COMSOL Multiphysics. A
single unit cell was modeled; periodic boundary conditions (BCs) in the $y$-direction
(corresponding to the normal incidence) and radiative BCs in the $x$-direction were assumed in the
simulation. The magnetic field distribution (color) and electric field vectors (arrows) of the most
dominant {\it magnetic resonance} mode is shown in Fig.~\ref{fig:experiment}(b). Both the strips
and the ground plate are gold with dielectric permittivity given by the Drude
model~\cite{Dolling_science06}: $\epsilon_{Au} = 1-\omega_p^2/[\omega(\omega+i\gamma)]$, where
$\omega_p=1.32{\times}10^{16}~s^{-1}$ and $\gamma=1.2{\times}10^{14}~s^{-1}$. By computing the
Ohmic losses and the power out-flux proportional to the Poynting vector, we have calculated $W_Q$
and $W_R$. These quantities were used in computing the resistive and radiative damping rates
$\omega_{iQ}=\omega_i \times W_Q/(W_Q+W_R)$ and $\omega_{iR}=\omega_i \times W_R/(W_Q+W_R)$,
respectively, which are required for calculating $z(\omega)$ from Eq.~(\ref{eq:impedance}). The
dependence of $\omega_{iQ}$ and $\omega_{iR}$ on the geometric parameters of the absorber are shown
in Fig.~\ref{fig:cross-section}.


According to Fig.~\ref{fig:cross-section}(a), the radiative loss associated with the eigenmode
dominates over the resistive loss whenever plasmonic strips are separated from the ground plate by
a large distance $G$. Therefore, $|r(\omega)|$ of an external beam tuned to $\omega=\omega_0$
($\omega_0(G)$ is plotted in the inset) is large, and the absorbance is small, which is not
surprising because the ground plate is highly reflective. On the other hand, if the strips
are close to the plate, a strong image dipole moment is generated behind the ground plate as shown
in Fig.~\ref{fig:schematic}(b). The resulting strong magnetic moment $\vec{m}$ directed along the
strips produces back-scattered fields with phase opposite to those produced by the reflection from
the ground plate. Such destructive interference reduces the total backward scattering. With a
suitable choice of $G$, critical coupling can be achieved when the scattering cross-section becomes
equivalent to the resistive cross-section. Such cancellation does not occur for large $G$s because
electric field is mostly concentrated between the adjacent strips, not between the strips and the
ground plate. Therefore, the magnetic moment is too weak to significantly reduce ground plate
reflection. Other geometric parameters also affect the total radiative loss. For a fixed
periodicity $L=350$~nm, we observe from Fig.~\ref{fig:cross-section} that radiative loss can be
decreased by increasing the strip width $W$ or decreasing the strip thickness $D$.

Assuming the single dominant resonance of the absorber, the peak
reflectivity $|r(\omega)|^2 = |(z-1)/(z+1)|^2$ at the resonant
frequency $\omega=\omega_0$ can be recast as
\begin{equation}\label{eq:reflectivity}
    |r(\omega_0)|^2 =
    \left|
    \frac{\omega_{iQ}-\omega_{iR}}{\omega_{iQ}+\omega_{iR}}
    \right|^2,
\end{equation}
confirming that ``perfect'' peak absorption is achieved if $\omega_{iR}=\omega_{iQ}$. The peak
absorbance $A=1-|r(\omega_0)|^2$ obtained from Eq.~(\ref{eq:reflectivity}) is plotted in
Fig.~\ref{fig:cross-section}(d) (markers) and compared with the peak absorbance obtained from
driven simulations (solid lines) for varying structure parameters. Excellent agreement is found
between the single-pole impedance approach based on eigenvalue simulations and the driven
simulations that include incident waves with frequencies scanned over a wide spectral range. The
advantage of the eigenvalue-based approach is that only a single simulation is required to compute
$z$ and, therefore, the peak absorbance. A full frequency scan (i.e., multiple FEM simulations) is
needed to find the peak absorbance using driven simulations.

One interesting consequence of Eq.~(\ref{eq:reflectivity}) is that the absorbance can still be
large even when the radiative loss is considerably mismatched from the resistive loss. For example,
when $\omega_{iR}=1.5~\omega_{iQ}$, the absorbance still reaches 96$\%$. We refer to this
phenomenon as the persistence of high absorptivity despite considerable variations of the
absorber's parameters. This insensitivity to the strip width $W$, strip thickness $D$, and
strip-to-ground plate spacing $G$ is shown in Fig.~\ref{fig:cross-section}(d): the absorbance stays
above $70\%$ despite considerable variation (by as much as a factor $4$) of these parameters.

To investigate the angular response of the plasmonic absorber, we have computed eigenmodes and
eigenfrequencies of ``leaky" surface plasmons propagating along the periodicity direction $y$ with
a finite wavenumber $k_y$. As long as $ck_y<\omega$, such plasmons are ``leaky" because they can
directly couple to the radiation with the frequency $\omega$ incident at the angle
$\theta=\arcsin(ck_y/\omega)$ with respect to the (normal) $x$-direction. Therefore, they can be
responsible for wide-angle absorption of the obliquely incident radiation. Just as for
the normal incidence, one expects critical coupling and ``perfect" absorption when $\omega_{iQ}
\approx \omega_{iR}$. Indeed, as Fig.~\ref{fig:angular_theory} indicates, the two damping rates are
very close to each other. Moreover, the resonant wavelength $\lambda_0=2\pi c/\omega_0$ remains
essentially flat for all values of $k_y$. The consequence of a flat response is persistently high
absorbance (between $80\%$ and $100\%$) of the plasmonic structure as the incidence angle $\theta$
is varied in the $0^\circ < \theta < 80^\circ$ range. In Section~\ref{sec:spectroscopy}, we present
experimental confirmation of wide-angle absorption.

\section{Fabrication of the absorber}\label{sec:fabrication}

In this section we provide a detailed technical description of the fabrication procedure of the
Large Area Wide-Angle Spectrally Selective Plasmonic Absorber (LAWASSPA). The structure was
fabricated using nano-imprint lithography~\cite{chou_jvac96,johnson_spie03}, which provides a
means to create many replicate copies of devices via pattern transfer from a quartz master template
that is initially defined using electron beam lithography (EBL). Once the EBL pattern is etched
into quartz, the patterned quartz template is used to stamp photoresist layers under UV exposure
(i.e., a step-and-flash process).  Release of the quartz template leaves the desired relief pattern
on the wafer surface. Subsequent dry etching transfers the pattern into the desired layer, which
are thin gold strips in this work.  The feature dimensions are limited only by the resolution of
the initial electron beam process, which is 50~nm for the JEOL JBX6000 EBL tool at the University
of Texas at Austin's Microelectronic Research Center (MRC). For volume production, resolution can
be as low as 20~nm~\cite{imprints}.

\begin{figure}[t]
\centering
\includegraphics[width=.48\textwidth]{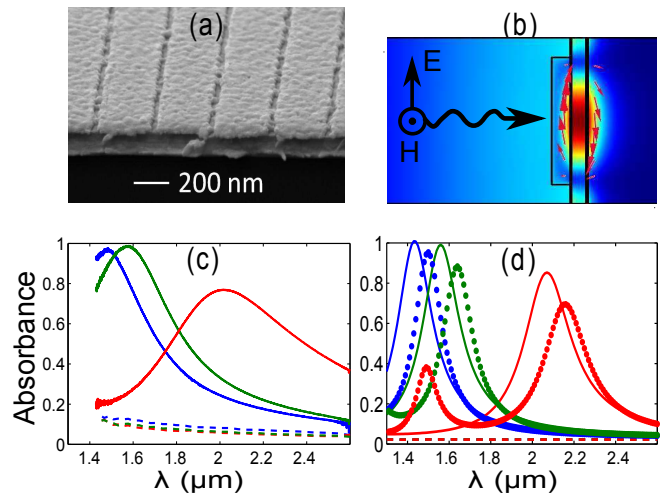} \
\caption{(a) SEM image of the strip absorber structure. (b) Simulated field profile at the
resonance. Color: $|H|$, arrow: {\bf E} field. (c) Measured and (d) simulated absorbance with
polarization perpendicular (solid lines) and parallel (dashed lined) to the strips. The dimensions
are [$L$,$W$]=[300~nm,230~nm] (blue), [330~nm,250~nm] (green), and [450~nm,350~nm] (red). $D$=30~nm
and $G$=22~nm for all three cases. The dotted lines in (d) are predicted absorbencies for square
patch arrays with identical parameters.} \label{fig:experiment}
\end{figure}

To fabricate the devices, Cr-Au-Cr layers (3~nm-94~nm-3~nm) were deposited on a 4\verb+"+ silicon
wafer by electron beam evaporation. A layer of  In$_2$O$_3$ (20~nm) followed by SiO$_2$ (1~nm) were
deposited by PECVD as the dielectric spacer. An additional Cr-Au layer (3~nm-30~nm) was
subsequently deposited on the dielectric spacer. Nano-imprint lithography is used to pattern the
top Cr-Au layer into metal strips, which begins by spin-coating a layer of $\rm {TranSpin}^{TM}$. A
number of combinations of metal strip width $W$ (100-300~nm) and pitch $L$ (175-450~nm) were
patterned into the quartz template as 1.4~mm$^2$ islands, providing a range of peak absorbance
wavelengths to investigate.

Nano-imprint lithography begins by dispensing curable liquid (silicon-based $\rm
{MonoMat}^{TM}$) with high precision over the desired area.  The quartz master is pressed against
the surface of the wafer (feature side down) and the $\rm {MonoMat}^{TM}$ flows to fill the relief
pattern in the quartz. Ultraviolet exposure sets the $\rm {MonoMat}^{TM}$ and the quartz master is
released from the surface leaving the imprinted pattern. After release, the residual $\rm
{MonoMat}^{TM}$ layer is dry etched by reactive ion etching (Oxford RIE, 200~V DC bias, 15 sccm
CHF$_3$, 7.5 sccm O$_2$, 25 mtorr) followed by dry etching of the $\rm {TranSpin}^{TM}$ layer
(200~V DC bias, 8 sccm O$_2$, 5 mtorr). The pattern is transferred into the Au/Cr layer using a
physical reactive ion etch in Ar (250~W, 50 sccm Ar, 40 mtorr). A final O$_2$ plasma etch (March
Asher) was used to remove remaining $\rm {MonoMat}^{TM}$ and $\rm {TranSpin}^{TM}$, yielding the
final device structure. An representative SEM image of the structure with $L$=300~nm and $W$=230~nm
is shown in Fig.~\ref{fig:experiment}(a). A single wafer contains a large number of LAWASSPA pixels
with different unit cell sizes. Each pixel has the size of approximately 1~mm$^2$, and is tuned to
a different resonant wavelength $\lambda_0 = 2\pi c/\omega_0$ determined by the unit cell's
dimensions $W$ and $L$.

\section{Angle-resolved Infrared Spectroscopy of Wide-Angle
Plasmonic Absorbers}\label{sec:spectroscopy}

In this section, we experimentally demonstrate that (i) the absorption peak of the infrared
plasmonic resonance can be spectrally tuned/controlled by changing the unit cell's dimensions as
described in Sec.~\ref{sec:theory}, and (ii) that the absorbance is wide-angle in agreement with
earlier theoretical calculations~\cite{wu_spie08,soukoulis_prb09,giessen_nl10,padilla_apl10}
and with the theory presented in Sec.~\ref{sec:theory}. To test (i), we have measured the
reflection spectra from different LAWASSPA pixels fabricated on a single wafer as
described in Sec.~\ref{sec:fabrication}. Reflectance ($R$) spectra were collected with a Thermo
Scientific Continuum microscope coupled to a Nicolet 6700 FTIR spectrometer using P-polarized
radiation and a 2~cm$^{-1}$ wavenumber resolution in the 7000~cm$^{-1}$--650~cm$^{-1}$
(1.4~$\mu$m--15.4~$\mu$m) spectral range. A wire-grid polarizer aligned the electric field
polarization along the periodicity direction $y$ as illustrated by Fig.~\ref{fig:experiment}(b).

Experimental absorbance for several LAWASSPA pixels of different dimensions are shown in
Fig.~\ref{fig:experiment}(c) by solid lines for the $y$-polarized infrared radiation. The incidence
angle of the infrared beam on the sample is approximately fixed at $25^\circ$ by microscope optics.
By changing the unit cell's dimensions $W$ and $L$ (see Fig.~\ref{fig:schematic} for definitions),
we have experimentally demonstrated spectral tunability of the plasmonic absorber between
$\lambda_0=1.6~\mu$m and $\lambda_0 = 2.0~\mu$m for the $y$-polarized beam. Much smaller absorption
measured for the orthogonal light polarization (electric field parallel to the long dimension $z$
of the strips) shown in Fig.~\ref{fig:experiment}(c) by the dashed lines does not exhibit any
spectral selectivity. As expected, the strip-based plasmonic absorber is polarization-sensitive: it
acts as a ``perfect" absorber/reflector for the $y$/$z$-polarized beams.

Simulations were performed and compared with experimental results. In calculations, the Drude model
for Au and the refractive index $n_{\rm spacer}=1.8$ for In$_2$O$_3$ were used. Theoretical
absorption spectra plotted in Fig.~\ref{fig:experiment}(d) for the same unit cell dimensions as the
fabricated samples (solid lines) show good agreement with the experimental spectra plotted in
Fig.~\ref{fig:experiment}(c). To account for thin post-etching metal residue between the strips
that was identified in SEM images, an extra 2~nm of Au on top of the spacer was assumed in
calculations. The effect of this metal residue is a slight blue-shifting of the resonance. While
the extra metal layer was not intentional in this set of experiments, it could become a useful
feature for specific applications that require electric contact between the metal strips of the
absorber.

Likewise, polarization selectivity of strip-based plasmonic absorbers could be useful for some
applications (e.g., thermal infrared emitters with controlled polarization state) and detrimental
for others (e.g., infrared photodetectors). Nevertheless, the results obtained for the strip-based
absorbers are directly relevant to patch-based~\cite{giessen_nl10,padilla_apl10}
polarization-insensitive absorbers shown in Fig.~\ref{fig:schematic}(b). Specifically, using COMSOL
simulations, we have found that the absorptivity of an array of {\it square} patches is essentially
identical to that of the infinite strips if the widths and periods ($W$ and $L$) of these
structures are chosen to be identical. The comparison between strip-based (solid lines) and
patch-based (dotted lines) plotted in Fig.~\ref{fig:experiment}(d) shows only a small systematic
red shift of the latter with respect to the former. This observation is useful from the design
standpoint because strip-based absorbers are much easier to simulate using two-dimensional FEM
simulations. Strip-based absorbers provide an excellent starting point for designing fully
three-dimensional patch-based plasmonic absorbers. We conclude that, while 2D strip arrays are
easier to analyze and engineer, absorbers with both strip and square patch arrays show the same
physics of critical coupling and high absorbance if their geometric parameters are near-identical.

\begin{figure}[t]
\centering
\includegraphics[width=.45\textwidth]{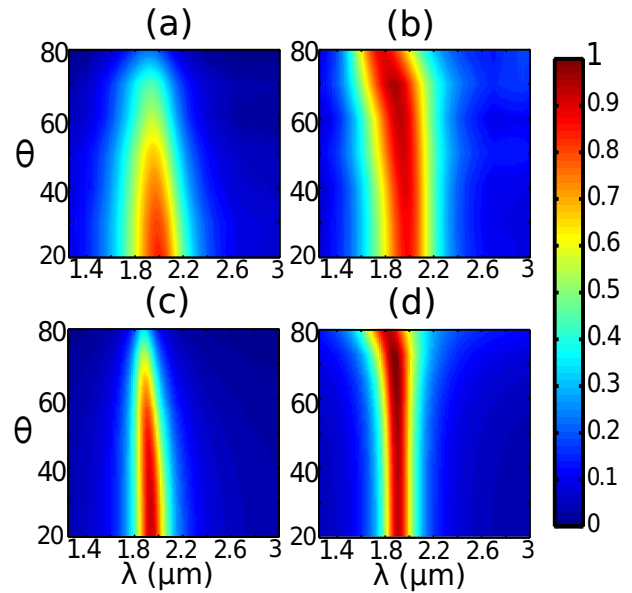} \
\caption{Angular-resolved absorption spectra $A(\theta,\lambda)$ for the strip-based plasmonic
absorber shown in Fig.~\ref{fig:experiment}. Experimental absorbance with (a) S- and (b)
P-polarized incidence (illustrated in Fig.~\ref{fig:schematic}(a)) is given in the upper panel, and
calculated absorbance with (c) S- and (d) P-polarized incidence is given in the lower
panel.}\label{fig:angular}
\end{figure}

If the structure is to be used as an effective absorber, the absorption rate must remain high for a
wide range of incident angles. Because the spectra shown in Fig.~\ref{fig:experiment}(c) were
obtained using a fixed-angle geometry (IR microscope), a separate optical setup enabling
variable-angle spectroscopy was used to test the wide-angle properties of the absorber. The
beamline is based on the FTIR spectrometer and uses standard infrared
focusing optics assembled on the optical table. It provides a means for variable incidence angles,
polarization control, and a 200-$\mu$m spot size. The geometric parameters of the strip-based
LAWASSPA are given by [$L$,$W$,$D$,$G$]=[300~nm,~230~nm,~10~nm,~17~nm], and the dielectric spacer
is SiO$_2$ ($n_{spacer}=1.5$). Angle-resolved absorption spectra were measured for both S and P
polarizations of the incident light as illustrated in Fig.~\ref{fig:schematic}(a), and presented as
a color map in Figs.~\ref{fig:angular}(a) and (b), respectively.

Both polarizations exhibit a fairly wide-angle absorptivity, although P polarization is much less
angle-dependent than S polarization. As was noted in earlier theoretical work~\cite{wu_spie08},
this dependence occurs because the plasmonic resonance responsible for ``perfect" absorption is primarily
magnetic and, therefore, is sensitive to the out-of-plane component of the magnetic field. The
qualitative agreement between theory (Fig.~\ref{fig:angular}(c,d)) and experiment
(Fig.~\ref{fig:angular}(a,b)) is excellent. Inhomogeneous broadening of the experimental spectra compared 
to theoretical predictions most likely results from
the absorber's structural imperfections. However, as we have demonstrated, the absorber design is robust against 
imperfections, and therefor the absorptivity remains high. One unintentional but interesting feature of this absorber
is that the absorption is not maximized at normal incidence. In fact, for P polarization, the
measured absorbance peaks at $A=96\%$ for $\theta=70^\circ$ and drops to $A=87\%$ for
$\theta=20^\circ$.

\section{Design of Small-Pixel Infrared Absorbers}\label{sec:small_pixel}

In this section, we discuss the implications of the wide-angle absorptivity of the infrared
plasmonic absorbers. One of the consequences of wide-angle absorptivity is that the coherence
propagation length $l_c$ of the surface plasmon responsible for the absorption is very short. If
$l_c \leq L$, then the interaction between the adjacent unit cells is small. Therefore, one can
envision combining a very small number of unit cells into a micro-pixel which can be as small as
one wavelength across. Such micro-pixels can act as an independent absorber/emitter. One possible
application of such micro-pixels could be a large focal plane array (FPA) for hyper-spectral
imaging. Such an array would consist of macro-pixels comprised of several micro-pixels, each of
which is tuned to a different wavelength. For example, a macro-pixel comprised of a $5 \times 5$
micro-pixels of width $w_j=\lambda_{0j}$ (where $1<j<25$) would be capable of monitoring 25
different wavelengths while remaining relatively small. The small size (several wavelengths across)
of a macro-pixel is essential for imaging/surveillance with a small angular resolution.

Our experimental data indeed indicates that $l_c < L$ and, therefore, the adjacent unit cell have
very little cross-talk between them. Quantitative estimate of the propagation length can be
obtained from the angular dependence of the absorptivity $A(\omega,k_y)$ which is plotted in
Fig.~\ref{fig:angular} as a function of the related variables $\lambda=2\pi c/\omega$ and $\theta$.
The propagation length along the $y$-direction is given by $l = v_g \tau$, where
$v_g=\partial\omega_0/\partial k_y$ is the group velocity in the $y$-direction, and $\tau$ is the
lifetime of the resonant mode that can be estimated as $\tau \approx A^{-1}
\partial A/\partial \omega$. Assuming that the absorption rate is
maximized at the angle-dependent eigenfrequency $\omega_0(k_y)$, and that the {\it peak} absorption
rate remains close to unity ($A(\omega_0,k_y) \approx 1$), we find that $|v_g| \approx \left|
\left( \partial A/\partial k_y \right)/\left( \partial A/\partial \omega \right) \right|$.
Combining these expressions for $\tau$ and $v_g$, and by assuming that $A \approx 1$, we obtain the
following simple estimate for the propagation length: $l_c \approx \left| \partial A/\partial k_y
\right|$. By noting that $ck_y = \omega \sin{\theta}$, we estimate the maximum plasmon propagation
length as $l_c < \left( \lambda/2\pi \right) \max\left[ \partial A/\partial \cos{\theta}\right]$. From the
experimental data for the wide-angle absorber, we estimate that the propagation length cannot
exceed $l_c < 300$~nm, i.e., about one period. Therefore, a resonant mode in a single cell is
significantly damped before reaching adjacent unit cells, and the cross-talk between them is small.

\begin{figure}[h]
\centering
\includegraphics[width=.45\textwidth]{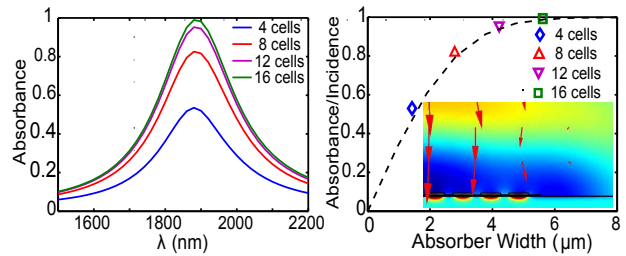}
\caption{(a) Absorbance by a finite number of unit cells illuminated by a Gaussian beam with an
intensity FWHM of 2.6~${\mu}$m. (b) Peak absorbance versus patterned-area width. Dashed line:
fractional overlap between the patterned area and the laser beam. Inset: a simulation of eight unit
cells (only half are shown) under the incoming Gaussian beam. Color: out-of-page magnetic field,
arrows: Poynting flux.}\label{fig:multicell}
\end{figure}

To substantiate our claim that the unit cells of the MM absorber function independently, the
absorbance of a limited number of MM unit cells is investigated theoretically. In
Fig.~\ref{fig:multicell}, we consider an incoming Gaussian beam with an intensity FWHM of
2.6~${\mu}$m impinging on MM surfaces consisting of 4, 8, 12, and 16 unit cells. The unit cells
possess [$L$,$W$,$D$,$G$]=[350~nm,~250~nm,~20~nm,~14~nm]. The metal is Au and the spacer is
SiO$_2$. Figure~\ref{fig:multicell}(a) shows absorbance with different numbers of unit cells,
whereas in Fig.~\ref{fig:multicell}(b) the peak absorbance is compared to the incoming flux within
the structured area. It is shown from the comparison that the MM absorber remains perfect absorbing
to the incident flux, even though the incoming field is not planar and the MM consists of as few as
four unit cells. This property can find applications in thermal imaging, where a pixel can be
defined by a small number of MM unit cells. On-chip spectrometers are also envisioned, in which
MM absorbers tuned to different wavelengths are patterned side-by-side on a single chip.

\section{Conclusions}\label{sec:conclusions}

In conclusion, we have demonstrated a simple design of an ultra-thin, wide-angle plasmonic absorber
exhibiting spectrally selective near-unity absorption. The high absorbance of the structure is
described in terms of critical coupling, where the scattering cross-section is equal to the
absorptive cross-section. The critical coupling perspective provides a guideline for designing MM
absorbers with tunable resonant frequencies: with low loss metals, the structure should be designed
in such a way that the resonances have low radiative loss. Based on this theory, the gold strips
and ground plate in this simple structure can be replaced by other metals, e.g., tungsten for
high-temperature applications, and the dielectric spacer layers can be replaced by absorbing
materials, e.g., semiconducting materials for photovoltaic applications. We have also developed a
convenient tool for designing such structures based on the eigenvalue simulations of the ``leaky"
plasmon resonances. These simulations are used for calculating the impedance of the absorbing
surface at arbitrary frequencies, which can then be used for predicting absorbance and reflectance.

Because a number of applications require inexpensive, large-area plasmonic absorbers, the
fabrication approach is paramount. We have fabricated plasmonic absorbers using step-and-flash
ultraviolet nano-imprint lithography, which has the advantages of reusable masks and large-area
fabrication compared with the serial fabrication methods. By patterning absorber pixels of various
dimensions over a large area, we have experimentally demonstrated that the spectral response of the
absorber can be controlled over a wide wavelength range ($1.5~\mu$m~$< \lambda < 2.0~\mu$m) while
maintaining a near-unity peak absorbance. Using angle-resolved infrared spectroscopy, we have
experimentally demonstrated that these Large-Area, Wide-Angle, Spectrally Selective Plasmonic
Absorbers (LAWASSPA) exhibit near-unity absorptivity for incidence angles from $20^\circ$ to
$80^\circ$. Moreover, we have experimentally demonstrated and theoretically validated that P-polarized light exhibits
wider-angle absorption than S-polarized light. Such wide-angle absorption indicates short
propagation lengths (less than one unit period) of the plasmonic mode, a feature useful in
designing ultra-small (one wavelength across) absorbing pixels that can be used as building
blocks for hyper-spectral focal plane arrays and infrared emitters.

\section{Acknowledgments}
This work is supported by the Office of Naval Research (ONR) grant
N00014-10-1-0929 and ONR's STTR program.

\bibliography{Absorber_Wu}

\begin{thebibliography}{28}
\expandafter\ifx\csname natexlab\endcsname\relax\def\natexlab#1{#1}\fi
\expandafter\ifx\csname bibnamefont\endcsname\relax
  \def\bibnamefont#1{#1}\fi
\expandafter\ifx\csname bibfnamefont\endcsname\relax
  \def\bibfnamefont#1{#1}\fi
\expandafter\ifx\csname citenamefont\endcsname\relax
  \def\citenamefont#1{#1}\fi
\expandafter\ifx\csname url\endcsname\relax
  \def\url#1{\texttt{#1}}\fi
\expandafter\ifx\csname urlprefix\endcsname\relax\def\urlprefix{URL }\fi
\providecommand{\bibinfo}[2]{#2}
\providecommand{\eprint}[2][]{\url{#2}}

\bibitem[{\citenamefont{Veselago}(1968)}]{veselago_spu68}
\bibinfo{author}{\bibfnamefont{V.~G.} \bibnamefont{Veselago}},
  \bibinfo{journal}{Soviet Physics Uspekhi} \textbf{\bibinfo{volume}{10}},
  \bibinfo{pages}{509} (\bibinfo{year}{1968}).

\bibitem[{\citenamefont{Smith et~al.}(2000)\citenamefont{Smith, Padilla, Vier,
  {Nemat-Nasser}, and Schultz}}]{smith_prl00}
\bibinfo{author}{\bibfnamefont{D.~R.} \bibnamefont{Smith}},
  \bibinfo{author}{\bibfnamefont{W.~J.} \bibnamefont{Padilla}},
  \bibinfo{author}{\bibfnamefont{D.~C.} \bibnamefont{Vier}},
  \bibinfo{author}{\bibfnamefont{S.~C.} \bibnamefont{{Nemat-Nasser}}},
  \bibnamefont{and} \bibinfo{author}{\bibfnamefont{S.}~\bibnamefont{Schultz}},
  \bibinfo{journal}{Phys. Rev. Lett.} \textbf{\bibinfo{volume}{84}},
  \bibinfo{pages}{4184} (\bibinfo{year}{2000}).

\bibitem[{\citenamefont{Pendry}(2000)}]{pendry_prl00}
\bibinfo{author}{\bibfnamefont{J.~B.} \bibnamefont{Pendry}},
  \bibinfo{journal}{Phys. Rev. Lett.} \textbf{\bibinfo{volume}{85}},
  \bibinfo{pages}{3966} (\bibinfo{year}{2000}).

\bibitem[{\citenamefont{Schurig et~al.}(2006)\citenamefont{Schurig, Mock,
  Justice, Cummer, Pendry, Starr, and Smith}}]{schurig_science06}
\bibinfo{author}{\bibfnamefont{D.}~\bibnamefont{Schurig}},
  \bibinfo{author}{\bibfnamefont{J.~J.} \bibnamefont{Mock}},
  \bibinfo{author}{\bibfnamefont{B.~J.} \bibnamefont{Justice}},
  \bibinfo{author}{\bibfnamefont{S.~A.} \bibnamefont{Cummer}},
  \bibinfo{author}{\bibfnamefont{J.~B.} \bibnamefont{Pendry}},
  \bibinfo{author}{\bibfnamefont{A.~F.} \bibnamefont{Starr}}, \bibnamefont{and}
  \bibinfo{author}{\bibfnamefont{D.~R.} \bibnamefont{Smith}},
  \bibinfo{journal}{Science} \textbf{\bibinfo{volume}{314}},
  \bibinfo{pages}{977} (\bibinfo{year}{2006}).

\bibitem[{\citenamefont{Alu and Engheta}(2004)}]{alu_ieee04}
\bibinfo{author}{\bibfnamefont{A.}~\bibnamefont{Alu}} \bibnamefont{and}
  \bibinfo{author}{\bibfnamefont{N.}~\bibnamefont{Engheta}},
  \bibinfo{journal}{IEEE Trans. Microw. Th. Tech.}
  \textbf{\bibinfo{volume}{52}}, \bibinfo{pages}{199} (\bibinfo{year}{2004}).

\bibitem[{\citenamefont{Chen et~al.}(2008)\citenamefont{Chen, O'Hara, Azad,
  Taylor, Averitt, Shrekenhamer, and Padilla}}]{chen_np08}
\bibinfo{author}{\bibfnamefont{H.-T.} \bibnamefont{Chen}},
  \bibinfo{author}{\bibfnamefont{J.~F.} \bibnamefont{O'Hara}},
  \bibinfo{author}{\bibfnamefont{A.~K.} \bibnamefont{Azad}},
  \bibinfo{author}{\bibfnamefont{A.~J.} \bibnamefont{Taylor}},
  \bibinfo{author}{\bibfnamefont{R.~D.} \bibnamefont{Averitt}},
  \bibinfo{author}{\bibfnamefont{D.~B.} \bibnamefont{Shrekenhamer}},
  \bibnamefont{and} \bibinfo{author}{\bibfnamefont{W.~J.}
  \bibnamefont{Padilla}}, \bibinfo{journal}{Nature Photon.}
  \textbf{\bibinfo{volume}{2}}, \bibinfo{pages}{295} (\bibinfo{year}{2008}).

\bibitem[{\citenamefont{Landy et~al.}(2008{\natexlab{a}})\citenamefont{Landy,
  Sajuyigbe, Mock, Smith, and Padilla}}]{padilla_absorber_08}
\bibinfo{author}{\bibfnamefont{N.~I.} \bibnamefont{Landy}},
  \bibinfo{author}{\bibfnamefont{S.}~\bibnamefont{Sajuyigbe}},
  \bibinfo{author}{\bibfnamefont{J.~J.} \bibnamefont{Mock}},
  \bibinfo{author}{\bibfnamefont{D.~R.} \bibnamefont{Smith}}, \bibnamefont{and}
  \bibinfo{author}{\bibfnamefont{W.~J.} \bibnamefont{Padilla}},
  \bibinfo{journal}{Phys.~Rev.~Lett.~} \textbf{\bibinfo{volume}{100}},
  \bibinfo{pages}{207402} (\bibinfo{year}{2008}{\natexlab{a}}).

\bibitem[{\citenamefont{Coutts}(1999)}]{coutts_rsev99}
\bibinfo{author}{\bibfnamefont{T.~J.} \bibnamefont{Coutts}},
  \bibinfo{journal}{Renewable and Sustainable Energy Rev.}
  \textbf{\bibinfo{volume}{3}}, \bibinfo{pages}{77} (\bibinfo{year}{1999}).

\bibitem[{\citenamefont{Laroche et~al.}(2006)\citenamefont{Laroche, Carminati,
  and Greffet}}]{laroche_jap06}
\bibinfo{author}{\bibfnamefont{M.}~\bibnamefont{Laroche}},
  \bibinfo{author}{\bibfnamefont{R.}~\bibnamefont{Carminati}},
  \bibnamefont{and} \bibinfo{author}{\bibfnamefont{J.~J.}
  \bibnamefont{Greffet}}, \bibinfo{journal}{J. Appl. Phys.}
  \textbf{\bibinfo{volume}{100}}, \bibinfo{pages}{063704}
  (\bibinfo{year}{2006}).

\bibitem[{\citenamefont{Tvingstedt et~al.}(2007)\citenamefont{Tvingstedt,
  Persson, Inganas, Rahachou, and Zozoulenko}}]{tvingstedt_apl07}
\bibinfo{author}{\bibfnamefont{K.}~\bibnamefont{Tvingstedt}},
  \bibinfo{author}{\bibfnamefont{N.~K.} \bibnamefont{Persson}},
  \bibinfo{author}{\bibfnamefont{O.}~\bibnamefont{Inganas}},
  \bibinfo{author}{\bibfnamefont{A.}~\bibnamefont{Rahachou}}, \bibnamefont{and}
  \bibinfo{author}{\bibfnamefont{I.~V.} \bibnamefont{Zozoulenko}},
  \bibinfo{journal}{Appl. Phys. Lett.} \textbf{\bibinfo{volume}{91}},
  \bibinfo{pages}{113514} (\bibinfo{year}{2007}).

\bibitem[{\citenamefont{Avitzour
  et~al.}(2009{\natexlab{a}})\citenamefont{Avitzour, Urzhumov, and
  Shvets}}]{avitzour_prb09}
\bibinfo{author}{\bibfnamefont{Y.}~\bibnamefont{Avitzour}},
  \bibinfo{author}{\bibfnamefont{Y.~A.} \bibnamefont{Urzhumov}},
  \bibnamefont{and} \bibinfo{author}{\bibfnamefont{G.}~\bibnamefont{Shvets}},
  \bibinfo{journal}{Phys. Rev. B} \textbf{\bibinfo{volume}{79}},
  \bibinfo{pages}{045131} (\bibinfo{year}{2009}{\natexlab{a}}).

\bibitem[{\citenamefont{Wu et~al.}(2008)\citenamefont{Wu, Avitzour, and
  Shvets}}]{wu_spie08}
\bibinfo{author}{\bibfnamefont{C.}~\bibnamefont{Wu}},
  \bibinfo{author}{\bibfnamefont{Y.}~\bibnamefont{Avitzour}}, \bibnamefont{and}
  \bibinfo{author}{\bibfnamefont{G.}~\bibnamefont{Shvets}},
  \bibinfo{journal}{Proc. SPIE} \textbf{\bibinfo{volume}{7029}},
  \bibinfo{pages}{70290W} (\bibinfo{year}{2008}).

\bibitem[{\citenamefont{Diem et~al.}(2009)\citenamefont{Diem, Koschny, and
  Soukoulis}}]{soukoulis_prb09}
\bibinfo{author}{\bibfnamefont{M.}~\bibnamefont{Diem}},
  \bibinfo{author}{\bibfnamefont{T.}~\bibnamefont{Koschny}}, \bibnamefont{and}
  \bibinfo{author}{\bibfnamefont{C.~M.} \bibnamefont{Soukoulis}},
  \bibinfo{journal}{Phys. Rev. B} \textbf{\bibinfo{volume}{79}},
  \bibinfo{pages}{033101} (\bibinfo{year}{2009}).

\bibitem[{\citenamefont{Hao et~al.}(2010)\citenamefont{Hao, Wang, Liu, Padilla,
  Zhou, and Qiu}}]{padilla_apl10}
\bibinfo{author}{\bibfnamefont{J.}~\bibnamefont{Hao}},
  \bibinfo{author}{\bibfnamefont{J.}~\bibnamefont{Wang}},
  \bibinfo{author}{\bibfnamefont{X.}~\bibnamefont{Liu}},
  \bibinfo{author}{\bibfnamefont{W.~J.} \bibnamefont{Padilla}},
  \bibinfo{author}{\bibfnamefont{L.}~\bibnamefont{Zhou}}, \bibnamefont{and}
  \bibinfo{author}{\bibfnamefont{M.}~\bibnamefont{Qiu}},
  \bibinfo{journal}{Appl. Phys. Lett.} \textbf{\bibinfo{volume}{96}},
  \bibinfo{pages}{251104} (\bibinfo{year}{2010}).

\bibitem[{\citenamefont{Liu et~al.}(2010{\natexlab{a}})\citenamefont{Liu,
  Mesch, Weiss, Hentschel, and Giessen}}]{giessen_nl10}
\bibinfo{author}{\bibfnamefont{N.}~\bibnamefont{Liu}},
  \bibinfo{author}{\bibfnamefont{M.}~\bibnamefont{Mesch}},
  \bibinfo{author}{\bibfnamefont{T.}~\bibnamefont{Weiss}},
  \bibinfo{author}{\bibfnamefont{M.}~\bibnamefont{Hentschel}},
  \bibnamefont{and} \bibinfo{author}{\bibfnamefont{H.}~\bibnamefont{Giessen}},
  \bibinfo{journal}{Nano Letters} \textbf{\bibinfo{volume}{10}},
  \bibinfo{pages}{2342} (\bibinfo{year}{2010}{\natexlab{a}}).

\bibitem[{\citenamefont{Tao et~al.}(2008)\citenamefont{Tao, Bingham,
  Strikwerda, Pilon, Shrekenhamer, Landy, Fan, Zhang, Padilla, and
  Averitt}}]{tao_prb08}
\bibinfo{author}{\bibfnamefont{H.}~\bibnamefont{Tao}},
  \bibinfo{author}{\bibfnamefont{C.~M.} \bibnamefont{Bingham}},
  \bibinfo{author}{\bibfnamefont{A.~C.} \bibnamefont{Strikwerda}},
  \bibinfo{author}{\bibfnamefont{D.}~\bibnamefont{Pilon}},
  \bibinfo{author}{\bibfnamefont{D.}~\bibnamefont{Shrekenhamer}},
  \bibinfo{author}{\bibfnamefont{N.~I.} \bibnamefont{Landy}},
  \bibinfo{author}{\bibfnamefont{K.}~\bibnamefont{Fan}},
  \bibinfo{author}{\bibfnamefont{X.}~\bibnamefont{Zhang}},
  \bibinfo{author}{\bibfnamefont{W.~J.} \bibnamefont{Padilla}},
  \bibnamefont{and} \bibinfo{author}{\bibfnamefont{R.~D.}
  \bibnamefont{Averitt}}, \bibinfo{journal}{Phys. Rev. B}
  \textbf{\bibinfo{volume}{78}}, \bibinfo{pages}{241103(R)}
  (\bibinfo{year}{2008}).

\bibitem[{\citenamefont{Urzhumov and Shvets}(2008)}]{urzh_shvets_ssc08}
\bibinfo{author}{\bibfnamefont{Y.}~\bibnamefont{Urzhumov}} \bibnamefont{and}
  \bibinfo{author}{\bibfnamefont{G.}~\bibnamefont{Shvets}},
  \bibinfo{journal}{Solid State Comm.} \textbf{\bibinfo{volume}{146}},
  \bibinfo{pages}{208} (\bibinfo{year}{2008}).

\bibitem[{\citenamefont{Johnson et~al.}(2003)\citenamefont{Johnson, Bailey,
  Dickey, Smith, Kim, Mancini, Dauksher, Nordquist, and
  Resnick}}]{johnson_spie03}
\bibinfo{author}{\bibfnamefont{S.}~\bibnamefont{Johnson}},
  \bibinfo{author}{\bibfnamefont{T.}~\bibnamefont{Bailey}},
  \bibinfo{author}{\bibfnamefont{M.}~\bibnamefont{Dickey}},
  \bibinfo{author}{\bibfnamefont{B.}~\bibnamefont{Smith}},
  \bibinfo{author}{\bibfnamefont{E.}~\bibnamefont{Kim}},
  \bibinfo{author}{\bibfnamefont{D.}~\bibnamefont{Mancini}},
  \bibinfo{author}{\bibfnamefont{W.}~\bibnamefont{Dauksher}},
  \bibinfo{author}{\bibfnamefont{K.}~\bibnamefont{Nordquist}},
  \bibnamefont{and} \bibinfo{author}{\bibfnamefont{D.}~\bibnamefont{Resnick}},
  \bibinfo{journal}{Proc. SPIE} \textbf{\bibinfo{volume}{5037}},
  \bibinfo{pages}{197} (\bibinfo{year}{2003}).

\bibitem[{\citenamefont{Avitzour
  et~al.}(2009{\natexlab{b}})\citenamefont{Avitzour, Urzhumov, and
  Shvets}}]{yoav_prb09}
\bibinfo{author}{\bibfnamefont{Y.}~\bibnamefont{Avitzour}},
  \bibinfo{author}{\bibfnamefont{Y.~A.} \bibnamefont{Urzhumov}},
  \bibnamefont{and} \bibinfo{author}{\bibfnamefont{G.}~\bibnamefont{Shvets}},
  \bibinfo{journal}{Phys. Rev. B} \textbf{\bibinfo{volume}{79}},
  \bibinfo{pages}{045131} (\bibinfo{year}{2009}{\natexlab{b}}).

\bibitem[{\citenamefont{Liu et~al.}(2010{\natexlab{b}})\citenamefont{Liu,
  Starr, Starr, and Padilla}}]{padilla_prl10}
\bibinfo{author}{\bibfnamefont{X.}~\bibnamefont{Liu}},
  \bibinfo{author}{\bibfnamefont{T.}~\bibnamefont{Starr}},
  \bibinfo{author}{\bibfnamefont{A.~F.} \bibnamefont{Starr}}, \bibnamefont{and}
  \bibinfo{author}{\bibfnamefont{W.~J.} \bibnamefont{Padilla}},
  \bibinfo{journal}{Phys. Rev. Lett.} \textbf{\bibinfo{volume}{104}},
  \bibinfo{pages}{207403} (\bibinfo{year}{2010}{\natexlab{b}}).

\bibitem[{\citenamefont{Landy et~al.}(2008{\natexlab{b}})\citenamefont{Landy,
  Sajuyigbe, Mock, Smith, and Padilla}}]{padilla_prl08}
\bibinfo{author}{\bibfnamefont{N.~I.} \bibnamefont{Landy}},
  \bibinfo{author}{\bibfnamefont{S.}~\bibnamefont{Sajuyigbe}},
  \bibinfo{author}{\bibfnamefont{J.~J.} \bibnamefont{Mock}},
  \bibinfo{author}{\bibfnamefont{D.~R.} \bibnamefont{Smith}}, \bibnamefont{and}
  \bibinfo{author}{\bibfnamefont{W.~J.} \bibnamefont{Padilla}},
  \bibinfo{journal}{Phys. Rev. Lett.} \textbf{\bibinfo{volume}{100}},
  \bibinfo{pages}{207402} (\bibinfo{year}{2008}{\natexlab{b}}).

\bibitem[{\citenamefont{Landy et~al.}(2009)\citenamefont{Landy, Bingham, Tyler,
  Jokerst, Smith, and Padilla}}]{padilla_prb09}
\bibinfo{author}{\bibfnamefont{N.~I.} \bibnamefont{Landy}},
  \bibinfo{author}{\bibfnamefont{C.~M.} \bibnamefont{Bingham}},
  \bibinfo{author}{\bibfnamefont{T.}~\bibnamefont{Tyler}},
  \bibinfo{author}{\bibfnamefont{N.}~\bibnamefont{Jokerst}},
  \bibinfo{author}{\bibfnamefont{D.~R.} \bibnamefont{Smith}}, \bibnamefont{and}
  \bibinfo{author}{\bibfnamefont{W.~J.} \bibnamefont{Padilla}},
  \bibinfo{journal}{Phys. Rev. B} \textbf{\bibinfo{volume}{79}},
  \bibinfo{pages}{125104} (\bibinfo{year}{2009}).

\bibitem[{\citenamefont{Smith et~al.}(2002)\citenamefont{Smith, Schultz,
  Markos, and Soukoulis}}]{smith_prb02}
\bibinfo{author}{\bibfnamefont{D.~R.} \bibnamefont{Smith}},
  \bibinfo{author}{\bibfnamefont{S.}~\bibnamefont{Schultz}},
  \bibinfo{author}{\bibfnamefont{P.}~\bibnamefont{Markos}}, \bibnamefont{and}
  \bibinfo{author}{\bibfnamefont{C.~M.} \bibnamefont{Soukoulis}},
  \bibinfo{journal}{Phys. Rev. B} \textbf{\bibinfo{volume}{65}},
  \bibinfo{pages}{195104} (\bibinfo{year}{2002}).

\bibitem[{\citenamefont{Haus}(1984)}]{haus_1984}
\bibinfo{author}{\bibfnamefont{H.~A.} \bibnamefont{Haus}},
  \emph{\bibinfo{title}{Waves and Fields in Optoelectronics}}
  (\bibinfo{publisher}{Prentice-Hall}, \bibinfo{year}{1984}).

\bibitem[{\citenamefont{{Neuner III} et~al.}(2009)\citenamefont{{Neuner III},
  Korobkin, Fietz, Carole, Ferro, and Shvets}}]{neuner_OL09}
\bibinfo{author}{\bibfnamefont{B.}~\bibnamefont{{Neuner III}}},
  \bibinfo{author}{\bibfnamefont{D.}~\bibnamefont{Korobkin}},
  \bibinfo{author}{\bibfnamefont{C.}~\bibnamefont{Fietz}},
  \bibinfo{author}{\bibfnamefont{D.}~\bibnamefont{Carole}},
  \bibinfo{author}{\bibfnamefont{G.}~\bibnamefont{Ferro}}, \bibnamefont{and}
  \bibinfo{author}{\bibfnamefont{G.}~\bibnamefont{Shvets}},
  \bibinfo{journal}{Opt. Lett.} \textbf{\bibinfo{volume}{34}},
  \bibinfo{pages}{2667} (\bibinfo{year}{2009}).

\bibitem[{\citenamefont{Dolling et~al.}(2006)\citenamefont{Dolling, Enkrich,
  Wegener, Soukoulis, and Linden}}]{Dolling_science06}
\bibinfo{author}{\bibfnamefont{G.}~\bibnamefont{Dolling}},
  \bibinfo{author}{\bibfnamefont{C.}~\bibnamefont{Enkrich}},
  \bibinfo{author}{\bibfnamefont{M.}~\bibnamefont{Wegener}},
  \bibinfo{author}{\bibfnamefont{C.~M.} \bibnamefont{Soukoulis}},
  \bibnamefont{and} \bibinfo{author}{\bibfnamefont{S.}~\bibnamefont{Linden}},
  \bibinfo{journal}{Science} \textbf{\bibinfo{volume}{312}},
  \bibinfo{pages}{892} (\bibinfo{year}{2006}).

\bibitem[{\citenamefont{Chou et~al.}(1996)\citenamefont{Chou, Krauss, and
  Renstrom}}]{chou_jvac96}
\bibinfo{author}{\bibfnamefont{S.~Y.} \bibnamefont{Chou}},
  \bibinfo{author}{\bibfnamefont{P.~R.} \bibnamefont{Krauss}},
  \bibnamefont{and} \bibinfo{author}{\bibfnamefont{P.~J.}
  \bibnamefont{Renstrom}}, \bibinfo{journal}{J. Vac. Sci. Technol. B}
  \textbf{\bibinfo{volume}{14}}, \bibinfo{pages}{4129} (\bibinfo{year}{1996}).

\bibitem[{imp()}]{imprints}
\bibinfo{howpublished}{\url{http://www.molecularimprints.com}}.

\end{thebibliography}

\end{document}